\documentclass[10pt,letterpaper]{article}
\usepackage[top=0.85in,left=2.75in,footskip=0.75in]{geometry}

\usepackage{amsmath,amssymb}

\usepackage{changepage}

\usepackage[utf8x]{inputenc}

\usepackage{textcomp,marvosym}

\usepackage{cite}

\usepackage{nameref,hyperref}

\usepackage[right]{lineno}

\usepackage{microtype}
\DisableLigatures[f]{encoding = *, family = * }

\usepackage[table]{xcolor}

\usepackage{array}

\usepackage{multirow}
\usepackage{array}
\usepackage{algorithm}
\usepackage{algpseudocode}

\newcolumntype{+}{!{\vrule width 2pt}}

\newlength\savedwidth

\newcommand\thickhline{\noalign{\global\savedwidth\arrayrulewidth\global\arrayrulewidth 2pt}%
\hline
\noalign{\global\arrayrulewidth\savedwidth}}

\raggedright
\usepackage[aboveskip=1pt,labelfont=bf,labelsep=period,justification=raggedright,singlelinecheck=off]{caption}

\makeatletter
\renewcommand{\@biblabel}[1]{\quad#1.}
\makeatother

\usepackage{lastpage,fancyhdr,graphicx}
\usepackage{epstopdf}
\fancyheadoffset[L]{2.25in}
\fancyfootoffset[L]{2.25in}
\begin{document}
\vspace*{0.2in}

\begin{flushleft}
{\Large
\textbf\newline{Symmetric core-cohesive blockmodel in preschool children's interaction networks} 
}
\newline
\\
Marjan Cugmas\textsuperscript{1},
Dawn DeLay\textsuperscript{2},
Ale{\v{s}} {\v{Z}}iberna\textsuperscript{1}
Anu{\v{s}}ka Ferligoj\textsuperscript{1,3}
\\
\bigskip
\textbf{1} Centre for Methodology and Informatics, Faculty of Social Sciences, University of Ljubljana, Ljubljana, Slovenia
\\
\textbf{2} Sanford School of Social and Family Dynamics, Arizona State University, Tempe, Arizona, USA 
\\
\textbf{3} International Laboratory for Applied Network Research, National Research University Higher School of Economics, Moscow, Russia
\\
\bigskip

*marjan.cugmas@fdv.uni-lj.si

\end{flushleft}
\section*{Abstract}
Researchers have extensively studied the social mechanisms that drive the formation of networks observed among preschool children. However, less attention has been given to global network structures in terms of blockmodels. A blockmodel is a network where the nodes are groups of equivalent units (according to links to others) from a studied network.  Cugmas et al. \cite{cugmas_mechanisms_2019} showed that mutuality, popularity, assortativity, and different types of transitivity mechanisms can lead the global network structure to the proposed asymmetric core-cohesive blockmodel. Yet, they did not provide any evidence that such a global network structure actually appears in any empirical data. In this paper, the symmetric version of the core-cohesive blockmodel type is proposed. This blockmodel type consists of three or more groups of units. The units from each group are internally well linked to each other while those from different groups are not linked to each other. This is true for all groups, except one in which the units have mutual links to all other units in the network. In this study, it is shown that the proposed blockmodel type appears in empirical interactional networks collected among preschool children. Monte Carlo simulations confirm that the most often studied social network mechanisms can lead the global network structure to the proposed symmetric blockmodel type. The units' attributes are not considered in this study.

\section*{Introduction}
One of the key attempts in sociology, and also in psychology, is to reveal the (social) mechanisms that are responsible for a given (social) output. When the relationships among individuals are studied, the social output is a social network. In social network analysis, there are different approaches to study the underlying social mechanisms of a given network. The main focus of earlier studies was on social mechanisms in the context of empirical networks while less attention was paid to the social mechanisms in the context of specific global network structures. Therefore, the general objective of the current study is to identify fundamental social mechanisms that guide the formation of a global network structure.

In this study, the global network structure is narrowed to a structure with three or more groups. The units from the first group (called the core group) have symmetric links established with all units in the network, while the units from the other groups (called cohesive groups) are internally well linked. The units from different cohesive groups are not linked to each other. This global network structure (called symmetric core-cohesive blockmodel, described in more detail in subsection Global network structure) is proposed since it is a combination of cohesive and symmetric core-periphery global network structures and because these global network structures can arise from the well-known transitivity \cite{bianconi_triadic_2014, cugmas_mechanisms_2019} and popularity \cite{cugmas_mechanisms_2019} mechanisms. These two mechanisms were found to be present in the formation of many liking and friendship networks collected among preschoolers (see subsection Local mechanisms).

The assumption made in this study is that the proposed global network structure appears among preschool children. Entrance to preschool brings a set of peers together who were previously unknown to one another. This is rare in the natural world and, thus, the shift into preschool peer groups offers a unique opportunity to assess and understand the mechanisms behind peer group formation. Preschool entry is also distinct from other social network settings in that it offers a closed network space in which peers interact. Preschool also provides a unique developmental context in which children are motivated, perhaps for the first time, to form new and enduring social relationships with similar-age peers \cite{fabes_next_2004, martin_social_2005}.

This assumption (of the emergence of the proposed global network structure) is tested by using the blockmodeling approach \cite{doreian_generalized_2005} on the symmetrized networks previously analyzed by Schaefer et al. \cite{schaefer_fundamental_2010}. Their study's main focus was the network dynamics rather than the global network structure. They showed that the selected local network mechanisms are important in such networks and that the importance of different local network mechanisms change throughout the school year. 

Building on the assumption (tested in this paper) that the proposed global network structure emerges in interactional preschool networks, the following research question is posed: \textbf{can the proposed global network structure appear due to the selected local network mechanisms without considering the nodes' attributes?} Here, the same local network mechanisms are assumed as in the study of Schaefer et al. \cite{schaefer_fundamental_2010} and other previous studies on preschool network dynamics. The research question is addressed using Monte Carlo simulations, specifically, by applying the proposed model from the family of network evolution models.

The study is relevant since understanding of the local network mechanisms at play, in the context of global network structures, is important while studying real (empirically observed) networks. Namely, the proposed global network structure's emergence at preschools raises very important developmental questions, e.g., how children in the core group differ from children in cohesive groups and what are the implications (if any) for their further individual development? Should such a global network structure be encouraged or discouraged? Is this a period where scholars may be able to document the emergence of social cliques and associated social norms? Will some children be integrated into cohesive groups, while others are left with minimal peer affiliations in the global network \cite{brown_friendships_2003}? 

The paper is organized as follows: a new global network structure is formally defined and the local network mechanisms are proposed and described (section Global network structure). Next, the global network structures of the empirical interactional preschool networks are analyzed (section The empirical case). The main research question concerned with the proposed global network structure's emergence is addressed in section Simulation approach and some conclusions are outlined.

\section*{Global network structure and local network mechanisms through structural processes}

In the following section, the proposed global network structure is defined in the blockmodel context. Different local network mechanisms that may drive the global network structure of preschool children's interactional networks towards the proposed one are discussed. 

\subsection*{Global network structure}

A blockmodel is a network in which the units are groups of equivalent units from a studied network \cite{doreian_generalized_2005}. The term reflects the fact that if a network is represented by a matrix, which is then split according to a partition (groups), blocks (submatrices) are formed in the matrix. The term ``block'' refers to a submatrix showing the links among units from the same or different group(s). Two selected units are structurally equivalent if they have the same pattern of links to the other units \cite{batagelj_direct_1992, lorrain_structural_1971}. The possible block types are identified through a selected definition of equivalence which is based on links among the units. Structural equivalence \cite{ lorrain_structural_1971} and its generalization, regular equivalence \cite{white_graph_1983}, are the most common. When structural equivalence is used, only null and complete blocks are possible. In ideal complete blocks, all possible links are present while no link exists in ideal null blocks.

\begin{figure}[!ht]
\begin{adjustwidth}{-2.25in}{0in} 
\caption{{\bf Example of an empirical network  and its blockmodeling solution}
(A) empirical network, (B) empirical network drawn in line with the blockmodeling solution, (C) blockmodel, (D) blockmodeling solution with two inconsistencies}
\label{fig1}
\includegraphics[width = 1.4\textwidth]{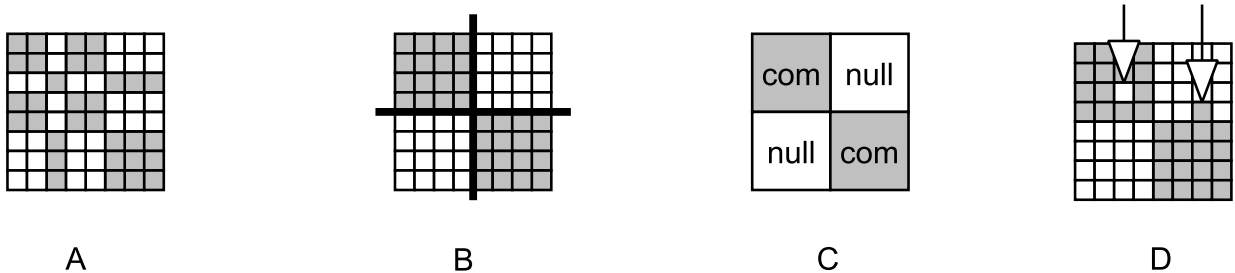}
\end{adjustwidth}
\end{figure}

A demonstration of blockmodeling according to structural equivalence is given in Fig~\ref{fig1}. The original network is visualized in matrix form in Fig~\ref{fig1}A. Here, each row and column represents a unit. Gray colored cells in a matrix represent a link from the $i$-th unit (row) to the $j$-th unit (column). Cells on the diagonal represent loops (a given unit is linked to itself). The units are permuted (see Fig~\ref{fig1}B) in such a way that those with the same pattern of links are placed together and form a cluster (group). Two groups are shown in Fig~\ref{fig1}B.

In the blockmodeling context, the clusters of units are shrinked into nodes. The blockmodel that is obtained is visualized in Fig~\ref{fig1}C. The obtained blockmodel has two nodes (shrinked groups). Here, two types of blocks appear (complete and null). Complete blocks are on the diagonal of the matrix because the units from both groups are internally linked to each other. Off-diagonal blocks refer to the relationships between different groups. Since the units from different groups are not linked to each other, the off-diagonal blocks are null blocks. 

The example represents an ideal case, meaning that there are all possible links in complete blocks and there is no link in the null blocks. However, this is unrealistic for empirical networks. In such networks, there are usually some non-links in complete blocks and some links in null blocks (see Fig~\ref{fig1}D). Such links are called errors or inconsistencies.

\begin{figure}[!ht]
\begin{adjustwidth}{-2.25in}{0in} 
\caption{{\bf Different representations of networks with a cohesive blockmodel, symmetric core-periphery blockmodel, and symmetric  core-cohesive blockmodel}
(A) cohesive blockmodel, (B) symmetric core-periphery blockmodel, (C) symmetric core-cohesive blockmodel}
\label{fig2}
\includegraphics[width = 1.4\textwidth]{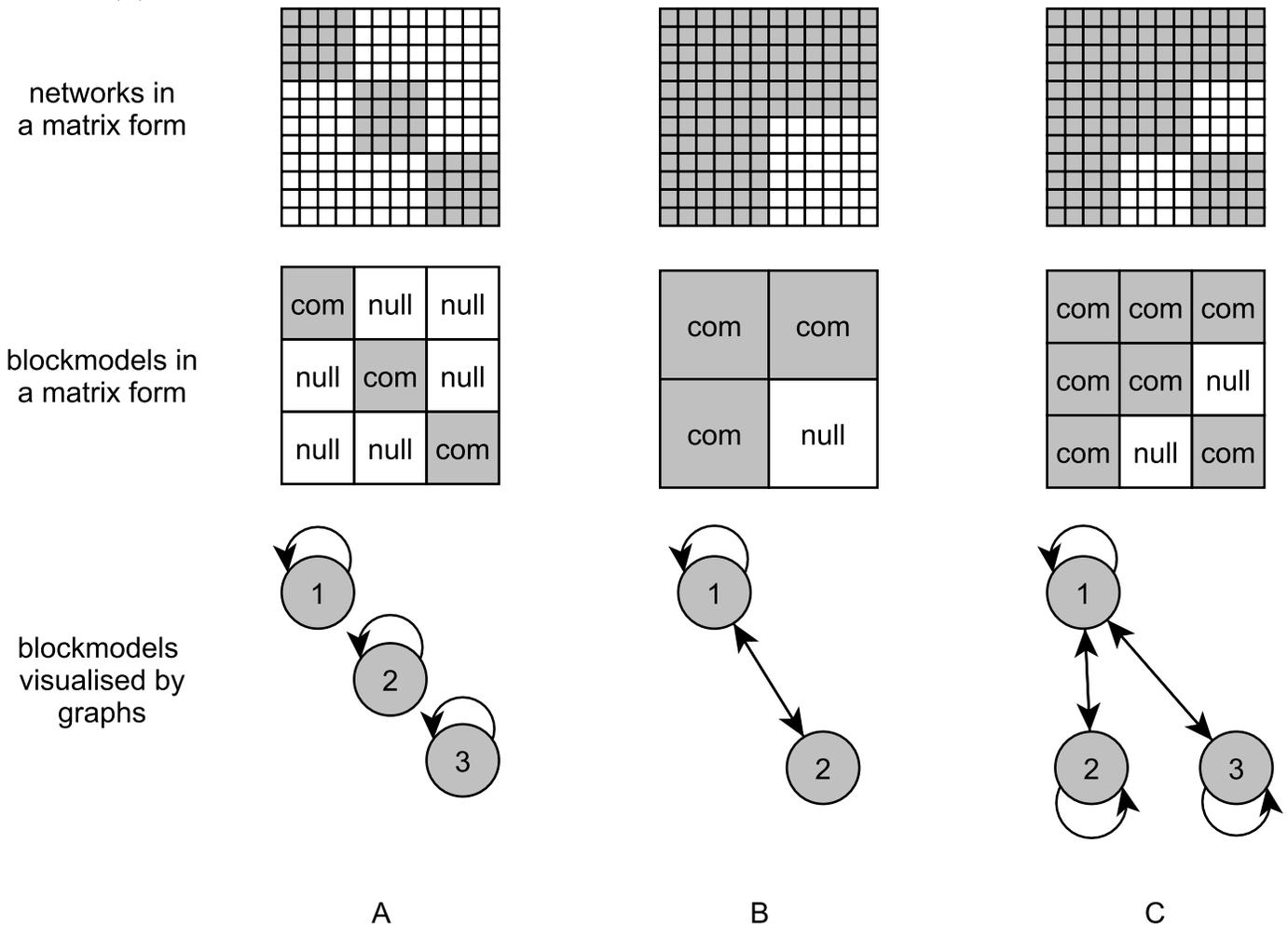}
\end{adjustwidth}
\end{figure}

There are several well-known blockmodel types, with two being the cohesive and (symmetric or asymmetric) core-periphery blockmodel types. The cohesive blockmodel type (Fig~\ref{fig2}A) contains at least two groups of units where units from different groups are not linked to each other, while all units inside each cohesive group are linked to each other. On the other hand, the symmetric core-periphery blockmodel (Fig~\ref{fig2}B) is defined by two groups of units. The units from the core group are internally well linked to each other and units from the periphery are not linked to each other. The units from the core are also linked to the units from the periphery and vice versa (in the asymmetric case, the units from the periphery are linked to the core ones or vice versa).

The newly proposed symmetric core-cohesive blockmodel type (Fig~\ref{fig2}C) is seen as a combination of a cohesive blockmodel and a symmetric core-periphery blockmodel. A symmetric core-cohesive blockmodel consists of one core group of units to which all units in the network are linked, and where units from the core group are linked to all other units in the network. The other units are classified into cohesive groups. Units from each cohesive group are internally linked to each other, while units from different cohesive groups are not linked to each other. The model can be extended in such a way that a group of units which are not linked to each other would also exist. 

\subsection*{Local mechanisms}

It has been hypothesized that the asymmetric core-cohesive blockmodel type might appear in networks observed among preschool children where the links are defined by ``friend nominations'' or ``liking" \cite{cugmas_mechanisms_2019}. Although research is lacking on the global network structure's evolution in the blockmodel context, many studies address the mechanisms that affect the creation and dissolution of ties. The social mechanisms of attraction most often discussed are mutuality (also known as reciprocity), popularity (also known as the Matthew effect or preferential attachment), transitivity, and assortativity (also known as assortative mixing or homophily) \cite{stadtfeld_emergence_2018}. The last one may be considered through the assortativity of in-degree  or other units' attributes, such as gender \cite{maccoby_gender_1987, vaughn_dyadic_2001}. Simulations confirm that an asymmetric core-cohesive blockmodel can appear as a result of the listed mechanisms. 

Since conducting longitudinal sociometric interviews with a high level of reliability and validity among preschool children might be too demanding for both the children and the researcher, the data analyzed in such settings are often observational. In such studies, a link is often operationalized as an interaction and therefore the observed links are undirected. If such interactions are considered as an indicator of friendship, popularity or liking, the same mechanisms must be considered when testing for the emergence of the symmetric core-cohesive blockmodel type. The following mechanisms are often discussed in the literature:

\begin{itemize}
\item \textbf{Mutuality} or reciprocity is defined through the reciprocation of ties and is one of the most fundamental social network mechanisms (besides creating links) and a basic feature of social life \cite{daniel_exponential_2013}. Analyzing 49- to 62-month-old preschool children, Snyder et al. \cite{snyder_social_1996} not only found that children spend much time with selected friends and less with others, but also strong evidence of mutuality. Observed mutual links in the empirical global network structures can also emerge since children prefer to interact with peers who are similar to themselves. This tendency often fosters the emergence of mutual peer relationships during childhood \cite{block_reciprocity_2015, kandel_homophily_1978, mcpherson_birds_2001, schaefer_fundamental_2010}.

The researcher cannot indirectly study this mutuality when analyzing non-directed interactional empirical networks. However, the mechanism can play a role in the process of creating the initiative for interactions (also see subsection The algorithm for generating networks).
\item \textbf{Popularity} is defined through an in-degree in social network analysis and is usually an operationalization of likeability or social status \cite{daniel_exponential_2013}. As a social network mechanism, popularity expresses the tendency to create links to others with a relatively high (in)degree. This is especially the case for less popular ones who wish to increase their own popularity by creating links with those who are most popular \cite{dijkstra_popularity_2013}. The fact that some units become more popular than others can relate to their personal attributes (e.g., wealth, being good at something, etc.) or positive or negative behavior \cite{cillessen_understanding_2005}.
\item \textbf{Transitivity} measures the tendency for triadic closure in networks - "the friends of my friends are also my friends". Transitivity in peer groups may arise from the increased propinquity of individuals who share mutual friends, or from a psychological need for balance - a convergence of third parties' evaluations \cite{schaefer_fundamental_2010}. 
\end{itemize}

Many empirical studies highlight the importance of these mechanisms. For example, Snyder et al. \cite{snyder_social_1996} noticed that children spend considerable time with selected friends and less with others. They also observed a strong mutual affiliation of friendships, which is subjected to the level of \emph{positive social consequences available from peers in the classroom}. 

Daniel et al. \cite{daniel_exponential_2013} used ERGM \cite{robins_introduction_2007} to study the mutuality, reciprocity, popularity, and transitivity mechanisms on the forming of affiliative ties in 19 Portuguese preschool peer groups. They found that all of these mechanisms are important for forming affiliative ties.

Schaefer et al. \cite{schaefer_fundamental_2010} studied the three most common network-formation mechanisms (reciprocity, popularity, and triadic closure) among preschool children throughout a school year in four waves using SIENA \cite{block_forms_2016, handcock_assessing_2003, snijders_introduction_2010}. They found the reciprocity effect is constant over time while the popularity effect is most important midway through the school year. The importance of the triadic closure effect increases over time, which is expected since very early on friendships are typically play-oriented dyads that primarily socialize children into group life \cite{hartup_friendships_1997}. When children gain more social contacts and greater confidence, they move into larger groups \cite{hartup_adolescents_1993}. 

\section*{The empirical case}

The hypothesis about the symmetric core-cohesive blockmodel being present in empirical interactional networks is tested in the subsections below. To this end, the empirical data collected among preschool children are analyzed using generalized blockmodeling. 

\subsection*{Data}

The data were collected as part of a bigger longitudinal study of young children's preparedness for school between 2004 and 2006 in Head Start preschools (the active consent to participate in the study was obtained from parents or guardians of children included in the study). The data were also analyzed in the study by Schaefer et al. \cite{schaefer_fundamental_2010}. The data are observational in nature, meaning that trained observers present in school classes recorded interactions among the children. Specifically, observers were present for several hours in a classroom two to three days per week. To ensure the greatest validity and reliability, two observers monitored the same children at the same time for 10 seconds. The order in which the observers watched over the children was random. When all children had been observed, the observers waited 5 minutes before repeating their observations (with a randomly reordered list of children). Children were observed in different activities, e.g. free play, talking, aggressive behavior, and others. The observers coded the type of activity in which a given child was involved and up to five other children with whom the selected child was interacting. Only the free-play data (data collected when children were able to play freely) are analyzed in this study. Children had to be observed at least 13 times during the whole school year to be included in the analysis. Based on the observational data, four complete networks are generated for each class. Each network's construction is based on a two-month period, as presented in Table~\ref{table1}. The networks are in matrix form in which each row and each column represents a child. The number of a given child's (ego, in a given row) observed interactions with other children (alter, in a given column) is shown in the corresponding cells of the matrix. The obtained networks were transformed from directed to undirected and binarized: there is a link between two children if the number of observed interactions is higher than the median (of the number of interactions between all possible pairs in the network) divided by two.

\begin{table}[!ht]
\begin{adjustwidth}{-2.25in}{0in} 
\centering
\caption{
{\bf Some basic descriptive statistics for the undirected networks.}}
\begin{tabular}{|l+r|r|r|r+r|r|r|r+r+r|} \hline
\multirow{3}{*}{{\rotatebox[origin=c]{90}{CLASS ID}}} & \multicolumn{4}{c+}{No. of units}                                                                                                                                                                                                          & \multicolumn{4}{c+}{No. of observations}  & \multirow{3}{*}{\begin{tabular}[c]{@{}l@{}}Age span \\ in the last \\ period \\ (in months)\end{tabular}} & \multirow{3}{*}{\begin{tabular}[c]{@{}l@{}}Percentage of \\ males in the \\ last period\end{tabular}} \\ \cline{2-9}
    & TP 1    & TP 2   & TP 3  & TP 4     & TP 1     & TP 2    & TP 3   & TP 4    &   &  \\ \cline{2-9}
 & \begin{tabular}[c]{@{}l@{}}Sep-Oct\\   2004\end{tabular} & \begin{tabular}[c]{@{}l@{}}Nov-Dec\\   2004\end{tabular} & \begin{tabular}[c]{@{}l@{}}Feb-Mar\\   2005\end{tabular} & \begin{tabular}[c]{@{}l@{}}Apr-May\\   2005\end{tabular} & \begin{tabular}[c]{@{}l@{}}Sep-Oct\\   2004\end{tabular} & \begin{tabular}[c]{@{}l@{}}Nov-Dec\\   2004\end{tabular} & \begin{tabular}[c]{@{}l@{}}Feb-Mar\\   2005\end{tabular} & \begin{tabular}[c]{@{}l@{}}Apr-May\\   2005\end{tabular} &                                                          &                                                         \\ \thickhline
1                    & 21                                                       & 20                                                       & 20                                                       & 19                                                       & 814                                                      & 510                                                      & 484                                                      & 321                                                      & 42-58                                                    & 63                                                      \\ \hline
2                    & 17                                                       & 17                                                       & 15                                                       & 14                                                       & 57                                                       & 95                                                       & 236                                                      & 374                                                      & 48-59                                                    & 50                                                      \\ \hline
3                    & 16                                                       & 17                                                       & 14                                                       & 14                                                       & 75                                                       & 200                                                      & 190                                                      & 184                                                      & 50-58                                                    & 50                                                      \\ \hline
4                    & 17                                                       & 18                                                       & 18                                                       & 16                                                       & 104                                                      & 410                                                      & 525                                                      & 548                                                      & 49-55                                                    & 69                                                      \\ \hline
5                    & 17                                                       & 17                                                       & 14                                                       & 14                                                       & 280                                                      & 406                                                      & 862                                                      & 413                                                      & 37-57                                                    & 50                                                      \\ \hline
6                    & 15                                                       & 15                                                       & 14                                                       & 14                                                       & 202                                                      & 343                                                      & 1005                                                     & 510                                                      & 46-59                                                    & 43                                                      \\ \thickhline
                       & \begin{tabular}[c]{@{}l@{}}Sep-Oct\\   2005\end{tabular} & \begin{tabular}[c]{@{}l@{}}Nov-Dec\\   2005\end{tabular} & \begin{tabular}[c]{@{}l@{}}Feb-Mar\\   2006\end{tabular} & \begin{tabular}[c]{@{}l@{}}Apr-May\\   2006\end{tabular} & \begin{tabular}[c]{@{}l@{}}Sep-Oct\\   2005\end{tabular} & \begin{tabular}[c]{@{}l@{}}Nov-Dec\\   2005\end{tabular} & \begin{tabular}[c]{@{}l@{}}Feb-Mar\\   2006\end{tabular} & \begin{tabular}[c]{@{}l@{}}Apr-May\\   2006\end{tabular} &                                                          &                                                         \\ \thickhline
7                    & 21                                                       & 19                                                       & 17                                                       & 16                                                       & 594                                                      & 564                                                      & 196                                                      & 589                                                      & 46-60                                                    & 44                                                      \\ \hline
8                    & 18                                                       & 18                                                       & 16                                                       & 16                                                       & 396                                                      & 432                                                      & 273                                                      & 855                                                      & 43-58                                                    & 69                                                      \\ \hline
9                    & 18                                                       & 18                                                       & 16                                                       & 15                                                       & 663                                                      & 406                                                      & 368                                                      & 1237                                                     & 37-59                                                    & 40                                                      \\ \hline
10                   & 16                                                       & 15                                                       & 15                                                       & 14                                                       & 931                                                      & 496                                                      & 309                                                      & 1609                                                     & 39-60                                                    & 64                                                      \\ \hline
11                   & 15                                                       & 16                                                       & 15                                                       & 15                                                       & 172                                                      & 241                                                      & 395                                                      & 574                                                      & 48-60                                                    & 47     \\ \thickhline                                                
\end{tabular}
\label{table1}
\end{adjustwidth}
\end{table}

The number of children varies between 14 and 21 across all networks. In the last period, the children were aged between 37 and 60 months and the share of males varied between 43\% and 69\%.

\subsection*{Methodology}

Binarized networks are blockmodeled to evaluate the global network structure. Blockmodeling is a way of reducing a large, potentially incoherent network to a smaller, comprehensible, and interpretable structure \cite{doreian_generalized_2005}. In a blockmodeling procedure, a list of allowed and forbidden block types is given. Since structural equivalence is used, these block types are null and complete. In order to not constrain the blockmodeling procedure, the relationships between groups (image matrix) is not pre-specified. 

The blockmodeling was done using the "blockmodeling" package \cite{ziberna_generalized_2018} for the R programming language. The number of iterations in the blockmodeling was 500 and 3 clusters were set for all networks. 

\subsection*{Results: empirical blockmodels}

\begin{figure}[!ht]
\begin{adjustwidth}{-2.25in}{0in} 
\caption{{\bf Obtained blockmodel structures for each class (ID1 to ID11) and each time period.}
Undirected and binarized empirical networks are considered. The obtained symmetric core-cohesive blockmodels are presented in the frame matrices.}
\label{fig3}
\includegraphics[width = 1.4\textwidth]{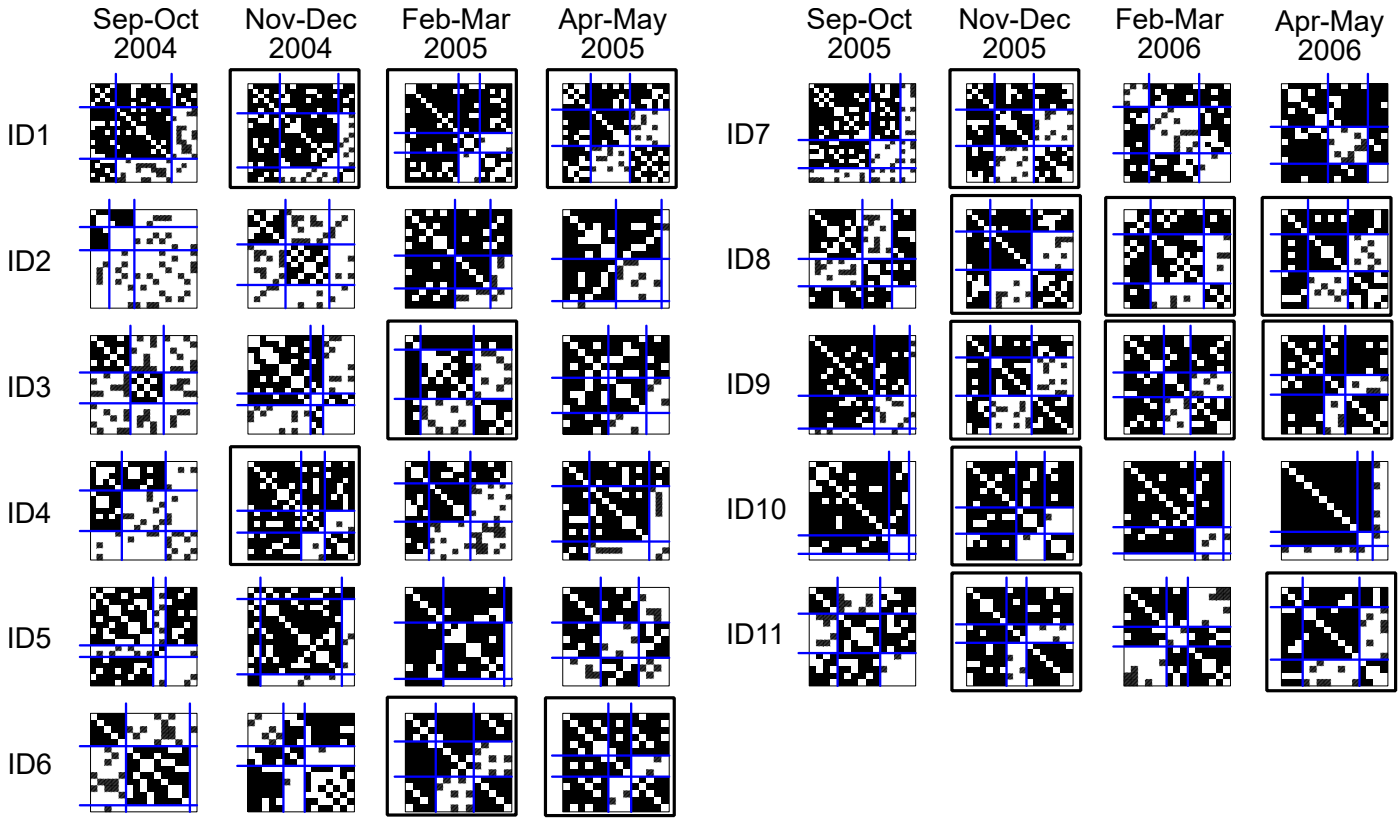}
\end{adjustwidth}
\end{figure}

Fig~\ref{fig3} gives the matrix representation of the analyzed networks. Each matrix corresponds to one network at a given time point. Black dots denote links. Children are ordered by rows and by columns in line with the solution from the blockmodeling. It can be seen that the networks are very dense, which is expected since interactional networks were observed in a closed environment (classroom). Some are almost complete. 

A symmetric core-periphery blockmodel structure (see the framed matrices) appears in almost all classes in at least one time period. It does not appear in just two classes (ID2 and ID5) out of 11 classes. In the other classes, the symmetric core-cohesive blockmodel appears in the 2nd time period (in 7 classes out of 11) or in the 3rd and 4th time periods (in 5 classes out of 11). The group sizes vary - in some cases, the core group consists of only 2 children (ID3 in Feb-Mar 2005) while in some other cases the core group consists of almost half the children (e.g., ID4 in Nov-Dec 2004 and ID43 in Nov-Dec 2005). 

Some of the blockmodels obtained are similar to the symmetric core-cohesive blockmodel type but are without links within the core (e.g., ID8 and ID11 in Sep-Oct 2005) or without one cohesive group (ID3 in Apr-May 2005).  

It has been shown that the proposed symmetric core-cohesive blockmodel type appears in empirical networks -- specifically, in interactional networks collected among preschool children. The question of whether the most commonly studied local network mechanisms can lead the global network structure towards the symmetric core-cohesive is addressed in the next section. Attributes of the units are not considered in this study. 

It should be noted that simulations can never prove that certain mechanisms cause global structure in empirical networks, only that they could cause it.

\section*{Simulation approach}

Cugmas et al. \cite{cugmas_mechanisms_2019}  have already shown that the mutuality, popularity, assortativity, and outgoing-two-path mechanisms  can lead towards the asymmetric core-cohesive blockmodel type and that different combinations of the local network mechanisms lead to this global network structure. The results were different when only one of the mechanisms was considered. For example, when only the popularity mechanism was considered, the resulting blockmodel was an asymmetric core-periphery blockmodel, while, on the other hand, the transitivity mechanism plays a role while forming the cohesive groups. Since the mechanisms are not independent, the role of the assortativity and mutuality mechanisms when considered together with popularity and transitivity in a blockmodel context is unclear and depends on the strengths of the other mechanisms.

In this paper, the \emph{symmetric} interactional networks of preschool children are studied. Therefore, the simulation approach proposed by Cugmas et al. \cite{cugmas_mechanisms_2019} for asymmetric networks is adapted to the symmetric case.

To evaluate whether the selected local network mechanisms can lead the global network structure towards the symmetric core-cohesive blockmodel, the adapted algorithm for generating networks is presented in the next subsection, and followed by definitions of the selected local network mechanisms. Many networks are generated using the proposed algorithm. In this generating process, different strengths of the mechanisms are considered. The global network structures of the generated networks are then evaluated in the Results subsection by applying the concepts of inconsistent blocks and relative fit value, which are also described in subsection Simulation design. 

\subsection*{The algorithm for generating networks}

A symmetric core-cohesive blockmodel may be generated in several ways by considering different local mechanisms. Two distinct approaches are identified with regard to whether symmetric or asymmetric links are generated:

\begin{itemize}
  \item \textbf{Symmetric (non)links:} here, it is assumed that all asymmetric links are reciprocated immediately. This means that a symmetric tie will exist if at least one of the actors chooses that tie and will not exist if at least one of the actors does not want it. The reciprocity mechanism is not considered in this case.
  \item \textbf{Asymmetric links:} only asymmetric links can be formed at a time. To achieve symmetric networks:
  \begin{itemize}
    \item the reciprocity mechanism must be considered. Here, a symmetric tie will exist if both actors choose the tie and will not exist if neither actor wants it (an asymmetric link will exist if only one chooses the tie). The generated networks can be asymmetric and therefore need to be analyzed as such or symmetrized before being further analyzed (e.g., by preserving all or only the symmetric links); and
    \item the reciprocity mechanism does not necessarily have to be considered, but the networks must be symmetrized before being further analyzed. This means that a symmetric tie will exist if at least one actor chooses the tie and will not exist if neither actor wants it.
  \end{itemize}
\end{itemize}

The observed interactional networks are symmetric by the definition of  ``interaction'', although the process which initiates interactions is asymmetric. In such a process, an ego has to initiate an interaction, while an alter can either: (i) accept (and reciprocate), (ii) tolerate, or (iii) reject (i.e., actively avoid) interaction. Even where an interaction is actively rejected by the alter, it can still be observed, although it is more likely to be recorded if it is either accepted or tolerated. Therefore, the approach where asymmetric ties are formed (by considering the mutuality mechanism) and the network is symmetrized, before being further analyzed, is the closest representation of the emergence of empirical networks.

Networks are represented in the form of an adjacency matrix $X$ of size $n*n$ where $n$ is the number of units. The possible values are 1s and 0s where 1s represent links, while 0s represent non-links. Because loops are not present, the diagonal values are 0. The proposed algorithm (see Algorithm~\ref{algorithm}) comes from the family of network evolution models (NEM) \cite{toivonen_comparative_2009} and can take initial networks with different blockmodels. 

\begin{algorithm}
\caption{The algorithm for generating networks used in this study}\label{algorithm}
\begin{algorithmic}[1]
    \Require initial network $X$
    \Require vector of strengths of the mechanisms $\theta$
    \Require probability of establishing a link $q$
    \Require number of iterations $k$
      \For{$l$ in $1:k$}
     	\State randomly select unit $i$ 
      	\State calculate network statistics according to the selected mechanisms for unit $i$ and all other units and save it in S
      	\State calculate $\phi = S\theta^T$
      	\State if $\phi \geq Q_3(\phi)$, classify unit $j$ into set $C$, where $Q_3$ is 3rd quartile
      	\State if $\phi \leq Q_1(\phi)$,  classify unit $j$ into set $F$, where $Q_1$ is 1st quartile
	\State with probability $q$ set $i \rightarrow j$ where $j$ is randomly selected from set $C$
	\State with probability $1-q$ set $i \not\rightarrow j$ where $j$ is randomly selected from set $F$
    \EndFor \\
    \Return generated network $X$
\end{algorithmic}
\end{algorithm}

The algorithm is iterational where the number of iterations $k$ can be determined based on the desired number of changes in the global network structure. Further, parameter $q$ must be set. It reflects the tendency towards the creation of a link and can be estimated based on the density of the network with the expected blockmodel. Yet, there is no guarantee the generated networks' density will equal $q$ since it depends on several factors, including the selected local mechanisms.

In the iterational process, a unit $i$ is randomly selected with probability $\frac{1}{n}$. Then, the network statistics $S$ are calculated based on the operationalized selected mechanisms (see the next subsection). These network statistics are weighted by the vector of strengths of local mechanisms $\theta$ producing vector $\phi = S\theta^T$. These units, for which it holds that their corresponding weighted network statistic is higher than or equal to the third quartile of all weighted network statistics, are classified in the set $C$ and are the candidates to accept the incoming tie from unit $i$. The other units, for which it holds that their corresponding weighted network statistic is lower than or equal to the first quartile of all weighted network statistics, are classified in the set $F$ and are candidates for being dissolved of an incoming tie by unit $i$. With probability $q$, the link from $i$ to randomly selected $j$ from set $C$ is set and with probability $1-q$ a non-link from $i$ to randomly selected $j$ from set $F$ is set. Since the unit can establish a link that already exists or dissolve a link that does not exist, there could be no visible change of a link upon a given iteration.

\subsection*{Formal definitions of the mechanisms}

The mechanisms are operationalized by different network statistics defined on a binary network, and normalized so that the minimum corresponding values are 0 and the maximum values are 1.

These network statistics ($S$) are weighted (by considering $\theta$) and summed to produce vector $\phi$ as described in the previous subsection. The local network mechanisms of which the network statistics are weighted with higher weights (in an absolute value) are more important in the network's evolution. The interpretation of a given mechanism depends on the sign of a corresponding weight.  For example, positively weighted popularity statistics refers to the tendency to create a link to those with a relatively high in-degree. On the contrary, a negative sign reflects the tendency to avoid establishing links to those with a relatively high in-degree.

\begin{figure}[!ht]
\caption{{\bf Illustrations of different mechanisms considered.}
(A) parameter $q$, (B) mutuality mechanism, (C) popularity mechanism, (D) assortativity mechanism, (E) transitivity mechanism, (F) outgoing-shared-partners mechanism}
\label{fig4}
\includegraphics[width = \textwidth]{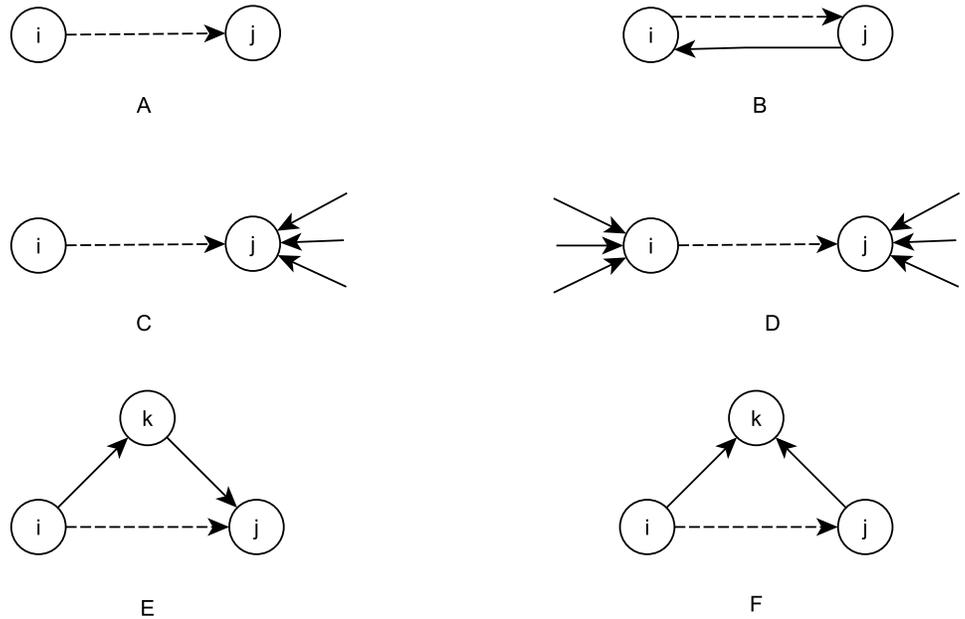}
\end{figure}

The mechanisms are defined in the same way as in a study of the asymmetric core-cohesive blockmodel type \cite{cugmas_mechanisms_2019}. Therefore, only a brief description of the proposed mechanisms is given here (the mechanisms are schematically shown in Fig~\ref{fig4}, where dashed lines illustrate the links under evaluation appear, are confirmed, or disappear):

\begin{enumerate}
\setcounter{enumi}{-1}
  \item \textbf{Parameter $q$} (Fig~\ref{fig4}A) reflects the tendency to have a link. Since this is not a focal mechanism, it is implemented in the NEM algorithm as parameter q and is therefore technically not considered as a mechanism in this study.
  \item The \textbf{mutuality mechanism} ($M$)  (Fig~\ref{fig4}B) reflects the tendency to reciprocate links.
  \item The \textbf{alter popularity mechanism} ($P$)  (Fig~\ref{fig4}C) reflects the tendency to create links to the most popular ones.
  \item 	The \textbf{assortativity mechanism} ($A$)  (Fig~\ref{fig4}D) reflects the tendency to create links to those units with the same level of popularity (in-degree). 
  \item 	The \textbf{transitivity mechanism} ($T$)  (Fig~\ref{fig4}E) is a tendency for a unit to directly connect to units, to which it is indirectly connected with (one or more) paths of length two (with more paths increasing the tendency).  
  \item 	The \textbf{outgoing-shared-partner mechanism} ($OSP$)  (Fig~\ref{fig4}F) represents a "structural homophily" effect which is traditionally based on similarity according to the units' attributes. In the case of the OSP, it is defined by similar choices of partners \cite{robins_closure_2009}.
\end{enumerate}

\subsection*{Simulation design}

The described NEM algorithm is used to generate the networks by considering the selected social mechanisms. Since different mechanism strengths are to be considered, 300 randomly selected $\theta$ are generated. The random values are generated by first sampling five values from the standard normal distribution $\Phi$ and then multiplying them by a scalar \cite{marsaglia_choosing_1972, muller_note_1959} (after such normalization, the sum of the squared elements of $\theta$ equals 1).

\begin{eqnarray}
\label{eq:theta}
	\theta = \frac{\Phi}{\sqrt{\sum \Phi^2_i}}
\end{eqnarray}

Within the NEM algorithm, parameter $q$ is set to 5/9 and a total of 116,490 iterations are applied. Parameter $q$, which indicates the tendency to create a link, is set arbitrarily but with reference to generating asymmetric core-cohesive blockmodels. Initial networks are empty with 24 units.

The 30 networks are generated for each $\theta$. Generalized blockmodeling for binary networks (on symmetrized generated networks, structural equivalence is used) is done after the selected number of iterations of the algorithm. More precisely, the intermediate number of iterations $m$, at which the global network structure is analyzed, is determined as $m_i=m_{i-1}*1.9$, where $m_1=100$ \cite{cugmas_mechanisms_2019}. This approach is used since most changes in the structure of the links happen at a lower number of iterations. 

Based on the generalized blockmodeling solution, the number of inconsistent blocks is calculated and used as the fit function. It is defined as the number of different blocks between the symmetric core-cohesive blockmodel with three groups and the empirically obtained blockmodel with three groups. 

Some $\theta$s that generate networks with the lowest number of inconsistent blocks are selected and further analyzed. For each network generated by the selected $\theta$s, the relative fit (RF) \cite{cugmas_mechanisms_2019} is calculated as

\begin{eqnarray}
\label{eq:RF}
	RF = 1 - \frac{P^m}{\frac{1}{k} \sum_{i=1}^{k} P^r_i}
\end{eqnarray}

\noindent where $P^m$ is the value of the criterion function \cite{batagelj_notes_1997, doreian_partitioning_1994} obtained on the empirical network and $P_i^r$ is the value of the criterion function obtained on the $i$-th randomized network. There are $k$ randomized networks. The mean value of the criterion function in the case of random networks is estimated by simulations. RF is a more detailed measure of the fit of a given blockmodel to the empirical data and its use is most valid when the presence of a given blockmodel type is confirmed by non-specified blockmodeling. Higher values indicate a better fit (the value of 1 indicates a perfect fit) and the expected value of the RF measure in the case of a random network is zero. 

\subsection*{Results: generated networks}

There are six different $\theta$s generating networks without any inconsistent block at the end of the iterations. Further, 76 different $\theta$s generate networks with the mean number of inconsistent blocks less than or equal to 0.5, and 109 different $\theta$s generate networks with the mean number of inconsistent blocks less than or equal to 1.

The $\theta$s that generate each network with a symmetric core-cohesive blockmodel are shown in Table~\ref{table2} along with the number of inconsistent blocks at a different number of iterations and the mean RF value of the generated networks. Although all the generated networks have the same blockmodel, they differ largely in the level of errors, expressed by RF.

\begin{table}[!ht]
\begin{adjustwidth}{-2.25in}{0in} 
\centering
\caption{
{\bf Mean number of inconsistent blocks and mean RF values with the corresponding parameter values.} For those $\theta$s which generated networks with the mean RF at the end of the iterations equal to zero. Initial is an empty network.}
\begin{tabular}{|l+l|r|r|r|r+r|r|l|l|l|l|l|l|l|l|l|l+l|}
\hline
\multirow{2}{*}{{\rotatebox[origin=c]{90}{$\theta$ ID}}} & \multicolumn{4}{c}{$\theta$}                          &      & \multicolumn{12}{c+}{NO. OF ITERATIONS}  & \multirow{2}{*}{} \\ \cline{2-18}
                             & {\rotatebox[origin=c]{90}{M}} & {\rotatebox[origin=c]{90}{P}} & {\rotatebox[origin=c]{90}{A}} & {\rotatebox[origin=c]{90}{T}} & {\rotatebox[origin=c]{90}{OSP}}  
                             & {\rotatebox[origin=c]{90}{100}}  & {\rotatebox[origin=c]{90}{190}}  & {\rotatebox[origin=c]{90}{361}}  & {\rotatebox[origin=c]{90}{686}}  
                             & {\rotatebox[origin=c]{90}{1,303}} & {\rotatebox[origin=c]{90}{2,478}} & {\rotatebox[origin=c]{90}{4,705}} & {\rotatebox[origin=c]{90}{8,939}} 
                             & {\rotatebox[origin=c]{90}{16,948}} & {\rotatebox[origin=c]{90}{32,969}} & {\rotatebox[origin=c]{90}{61,311}} & {\rotatebox[origin=c]{90}{116,490}} & {\rotatebox[origin=c]{90}{MRF}}        \\  \thickhline
136                          & -.18      & .74        & .37           & -.35         & .42  & 4.93 & 4.97 & 4.40 & 3.30 & 0.33  & 0.03  & 0.03  & 0.07  & 0.03   & 0.00   & 0.00   & 0.00    & 0.96              \\
25                           & -.43      & .27        & .66           & .25          & -.50 & 4.90 & 4.77 & 3.57 & 0.83 & 0.00  & 0.00  & 0.00  & 0.00  & 0.00   & 0.00   & 0.00   & 0.00    & 0.80              \\
279                          & .17       & -.11       & .43           & .60          & .65  & 4.73 & 4.93 & 3.43 & 0.20 & 0.00  & 0.07  & 0.00  & 0.00  & 0.00   & 0.00   & 0.00   & 0.00    & 0.50              \\
248                          & .11       & -.58       & .49           & .78          & -.38 & 4.73 & 4.97 & 4.00 & 1.37 & 0.33  & 0.00  & 0.00  & 0.03  & 0.00   & 0.00   & 0.00   & 0.00    & 0.48              \\
72                           & -.57      & .68        & .04           & -.46         & .10  & 4.90 & 4.97 & 4.13 & 3.77 & 0.70  & 0.20  & 0.03  & 0.00  & 0.10   & 0.00   & 0.00   & 0.00    & 0.35              \\
22                           & -.24      & -.51       & .21           & -.21         & -.78 & 5.00 & 5.10 & 4.45 & 2.03 & 0.17  & 0.03  & 0.00  & 0.00  & 0.00   & 0.00   & 0.00   & 0.00    & 0.26              \\ \thickhline
\end{tabular}
\label{table2}
\begin{flushleft} M = mutuality mechanism, P = popularity mechanism, A = assortativity mechanism, T = transitivity mechanism, OSP = outgoing-shared-partners mechanism, MRF = the mean RF value
\end{flushleft}
\end{adjustwidth}
\end{table}

\begin{figure}[!ht]
\begin{adjustwidth}{-2.25in}{0in} 
\caption{{\bf Mean RF for each blockmodel type visualized by lines and the distribution of the density visualized by boxplots.}
The networks are generated by considering $\theta=\{M=-0.18,P=0.74,A=0.37,T=-0.35,OSP=0.42\}$, $q=5/9$, $d_0=0$.}
\label{fig5}
\includegraphics[width = 1.4\textwidth]{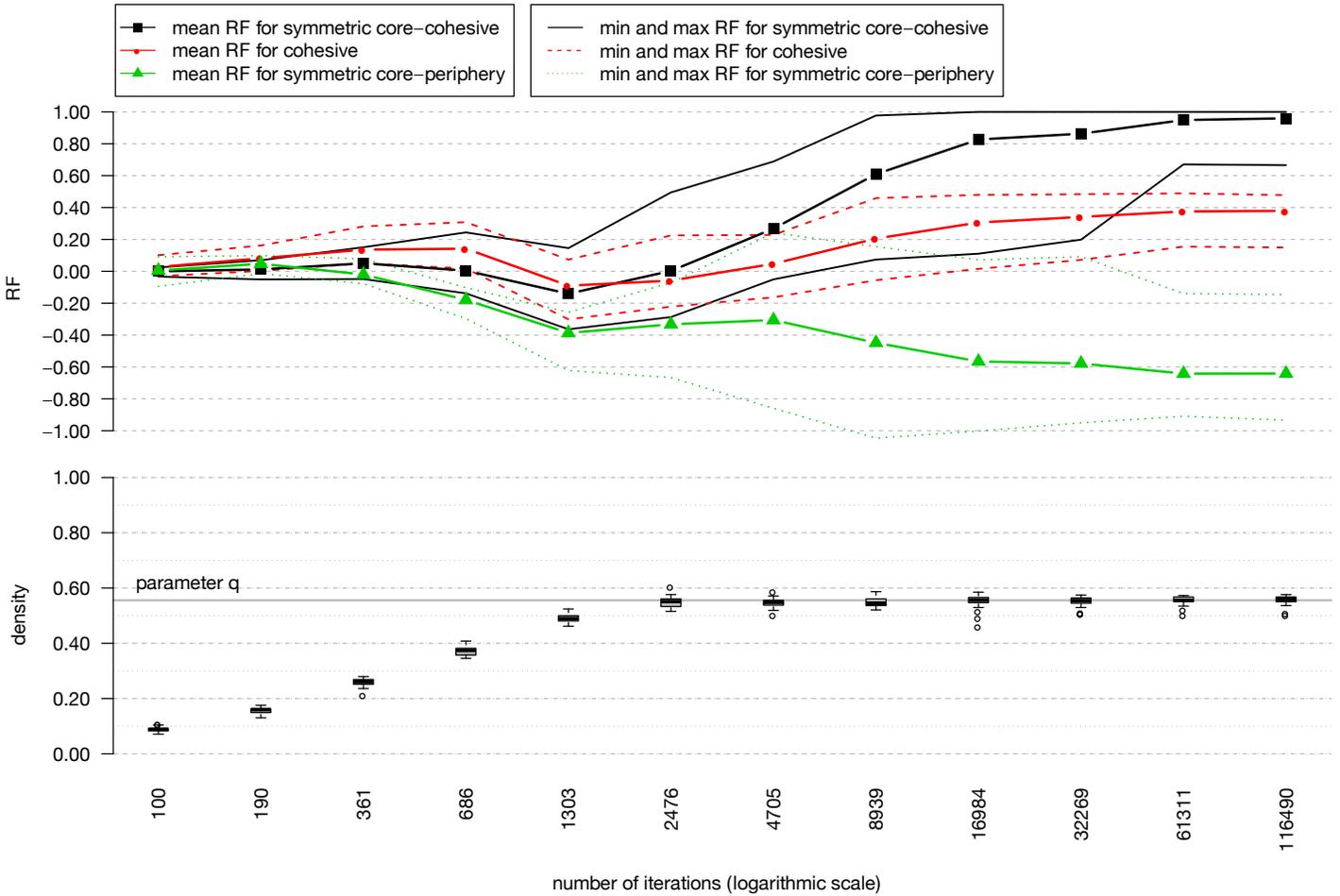}
\end{adjustwidth}
\end{figure}

\begin{figure}[!ht]
\begin{adjustwidth}{-2.25in}{0in} 
\caption{{\bf Some networks generated.}
The networks are generated by considering $\theta=\{M=-0.18,P=0.74,A=0.37,T=-0.35,OSP=0.42\}$, $q=5/9$, $d_0=0$. The networks are drawn in line with the blockmodels obtained by generalized blockmodeling (non-specified model). Networks for different repetitions of the algorithm for generating networks are drawn in lines for different numbers of iterations.
}
\label{fig6}
\includegraphics[width = 1.4\textwidth]{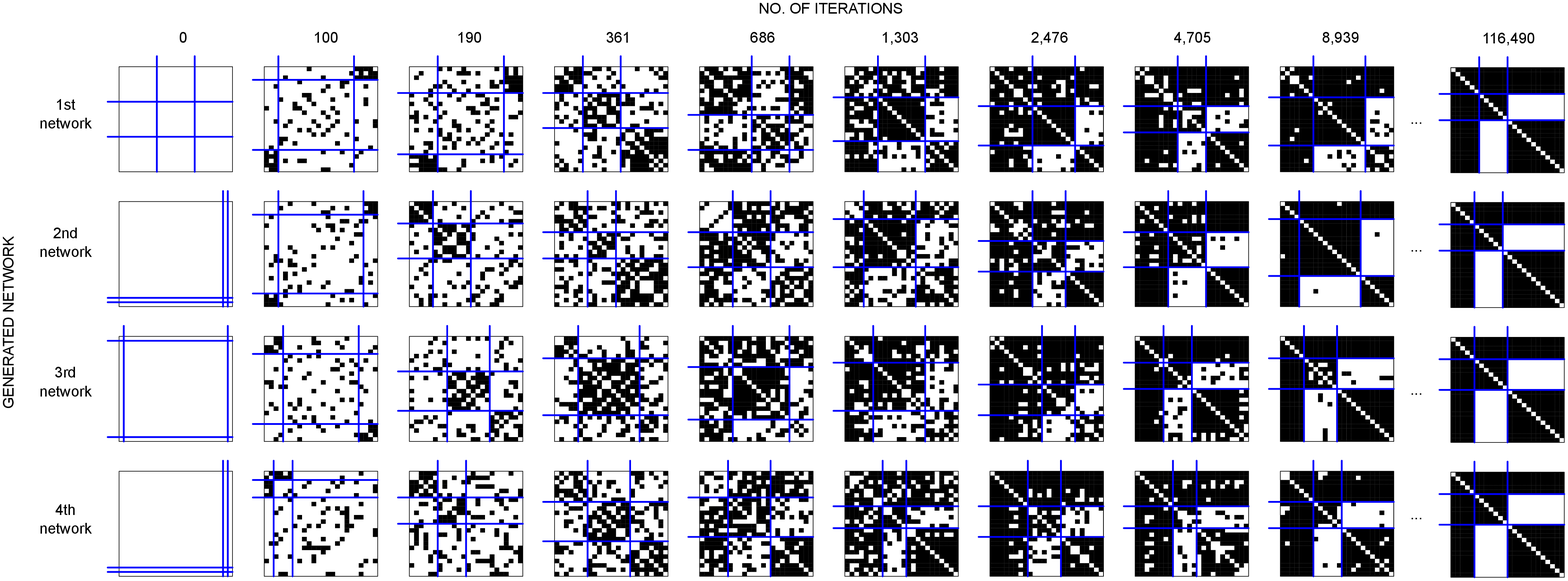}
\end{adjustwidth}
\end{figure}

A more detailed insight into RF for a selected $\theta$ is given in  Fig~\ref{fig5}. The mean RF values are calculated for the symmetric core-cohesive blockmodel type, cohesive blockmodel type, and symmetric core-periphery blockmodel type. All RF values are close to zero at the first 190 iterations. At such a low number of iterations, there are insufficient links to enable any of the considered blockmodel types to emerge. However, at 361 iterations, a global network structure, close to cohesive, can be visually recognized on the generated networks Fig~\ref{fig6}. Since there is a relatively high level of errors in null and complete blocks, the corresponding mean RF is very low. With a higher number of iterations (until 1,303 iterations), the mean RF, corresponding to all considered blockmodel types, is decreasing. At this step, the links among different groups are established yet, in some cases, links within the core units are not present. Moreover, there is a high level of errors in the null and complete blocks. After 1,303 iterations, the mean RF value for the core-cohesive and cohesive blockmodel is only increasing until 61,311 iterations. 

The mean RF, corresponding to the symmetric core-cohesive blockmodel type, is close to 1 at the end of the iterations, indicating the global network structure is the desirable one with almost no error in null and complete blocks (as confirmed in Fig~\ref{fig6}). The mean RF for the cohesive blockmodel is lower while the mean RF for the symmetric core-periphery blockmodel type is highly negative, indicating that the randomized networks fit this blockmodel type much more than the networks generated by using the proposed algorithm. 

\section*{Conclusion}

Interactional networks collected in preschool classrooms are studied in this paper. In such an environment, children start to form groups. Children within the individual groups spend more time with each other than they do with children from other groups. At the same time, a group of children is formed which spends a considerable amount of time with all the others from any group.

This leads to the newly proposed blockmodel type, i.e. symmetric core-cohesive. It consists of one group of units which are called core units and two or several other groups of units which are called cohesive groups. The units from all groups are internally linked to each other. The units from all cohesive groups are linked to the units from the core group and vice versa. The units from different cohesive groups are not linked to each other. 

The existence of this blockmodel type is evaluated on empirical data. The data were collected within a larger longitudinal study among preschool children in the United States between 2004 and 2006. The interactions among the children in classrooms were recorded and complete networks were formed. The symmetric core-cohesive blockmodel was found to be present in almost all analyzed classes in at least one time period. This proves that the proposed global structure (blockmodel type) is relevant for such data.

The most common local network mechanisms (popularity, assortativity, transitivity, and outgoing-shared-partners mechanism) are considered. Attributes of the units are not taken into account and the initial networks are empty. The adapted version of the algorithm proposed by Cugmas et al. \cite{cugmas_mechanisms_2019}  is used to generate the networks by considering the local network mechanisms. The results of the Monte Carlo simulations confirm that the selected mechanisms can generate networks with the symmetric core-cohesive blockmodel. The results do not imply that the global network structures of the empirical preschool networks collected in the 11 classes in the United States emerged due to the studied local network mechanisms. To address this question, a different methodology should be applied. 

The study is important in several ways given that understanding the emergence of peer network structure holds important implications for directing adaptive (prosocial) and redirecting maladaptive (bullying) peer network dynamics via intervention and prevention strategies. First, blockmodeling is shown to be an efficient way to describe and analyze empirical interactional network global structures. Second, understanding the link between the global network structure and the local network mechanisms in a given context is necessary for studying (e.g., modelling) the empirically obtained networks. It has been shown that the selected local network mechanisms are important in the formation of the symmetric core-cohesive blockmodel even without considering any further attributes of the units. 

\nolinenumbers


\end{document}